\documentclass[useAMS,usenatbib]{mn2e}
\usepackage{graphicx}

\title[Gemini GMOS--IFU spectroscopy of the compact H\,{\sc ii} galaxies Tol 0104-388 and 
Tol 2146-391]{Gemini GMOS--IFU spectroscopy of the compact H\,{\sc ii} galaxies Tol 0104-388 and 
Tol 2146-391: The dependence on the properties of the interstellar medium}
\author[P. Lagos et al.]{P. Lagos$^{1}$\thanks{E-mail: plagos@astro.up.pt}, 
E. Telles$^{2}$, A. Nigoche-Netro$^{3}$ and E. R. Carrasco$^{4}$\\
$^{1}$Centro de Astrof\'{\i}sica da Universidade do Porto, Rua das Estrelas, 4150-762 Porto, Portugal\\
$^{2}$Observat\'{o}rio Nacional, Rua Jos\'{e} Cristino, 77, Rio de Janeiro, 20921-400, Brazil\\
$^{3}$Instituto de Astronom\'{\i}a y Meteorolog\'{\i}a, Av, Vallarta 2602. Col. Arcos Vallarta. Guadalajara, Jalisco. C.P. 44130 M\'exico.\\
$^{4}$Gemini Observatory/AURA, Southern Operations Center, Casilla 603, La Serena, Chile}

\begin{document}


\pagerange{\pageref{firstpage}--\pageref{lastpage}} \pubyear{2012}

\maketitle

\label{firstpage}

\begin{abstract}
Using GMOS--IFU spectroscopic observations of the compact H\,{\sc ii}/BCD 
galaxies Tol 0104-388 and Tol 2146-391, we study the spatial distribution 
of emission lines, equivalent width EW(H$\beta$), extinction c(H$\beta$), ionization ratios 
([OIII] $\lambda$5007/H$\beta$, [SII] $\lambda\lambda$6717,6731/H$\alpha$ and [NII] $\lambda$6584/H$\alpha$), kinematics, 
and the chemical pattern (O/H, N/H and N/O) of the warm interstellar medium in these galaxies. 
We also investigate a possible dependence of these properties on the I(He \,{\sc ii} $\lambda$4686)/I(H$\beta$) ratio
and find no significant correlation between these variables. 
In fact, the oxygen abundances appear to be uniform
in the regions where the He \,{\sc ii} $\lambda$4686 emission line was measured. 
It can be interpreted in the sense that these correlations
are related to global properties of the galaxies and
not with small patches of the interstellar medium.
Although a possible weak N/H gradient is observed in Tol 2146-391, 
the available data suggest that the metals from previous star-formation 
events are well mixed and homogeneously distributed through the optical extent of these galaxies. 
The spatial constancy of the N/O ratio might be attributed to efficient transport 
and mixing of metals by starburst-driven super-shells,
powered by a plethora of unresolved star cluster in the inner part of the galaxies. 
This scenario agrees with the idea that most of the 
observed He \,{\sc ii} $\lambda$4686 emission line, in our sample of galaxies, 
is produced by radiative shocks.

\end{abstract}


\section{Introduction}

The ionizing radiation from newly formed stars and its interaction with 
the surrounding gas generate collisionally excited and recombination emission lines, 
that are commonly observed in starburst and low metallicity \citep[1/50Z$_{\odot}$-1/3Z$_{\odot}$;][]{KS83} 
dwarf galaxies, such as, H\,{\sc ii} or blue compact dwarf (BCD) galaxies. 
From recent studies it was concluded that the hardness of the ionizing radiation 
from the current star formation (SF) activity increases with decreasing metallicity \citep[e.g.,][]{G00,S03,TI05}
as is expected in primeval galaxies in the early universe. 
These first stars (population \,{\sc iii} stars) should be very massive and hot \citep[e.g.,][]{A02}, emitting
very hard ionizing radiation, thus very effective in ionizing hydrogen and helium, then strong 
He \,{\sc ii} emission lines are likely present in the spectra of these galaxies
\citep[e.g.,][]{S02,S03}. Originally, the low abundances of heavy
elements in HII/BCD galaxies and the non-detection of an old stellar population have given rise to the question
of whether they may be presently forming their first generation of stars \citep{SS70}.
Recent works, however, have shown that most H\,{\sc ii} galaxies seem to present an
underlying population of old stars from previous episodes of SF \citep[e.g.,][]{P96,TT97,C03}
suggesting an intermittent SF history with short intense SF episodes followed by long quiescent phases.

\begin{figure}
\includegraphics[width=70mm]{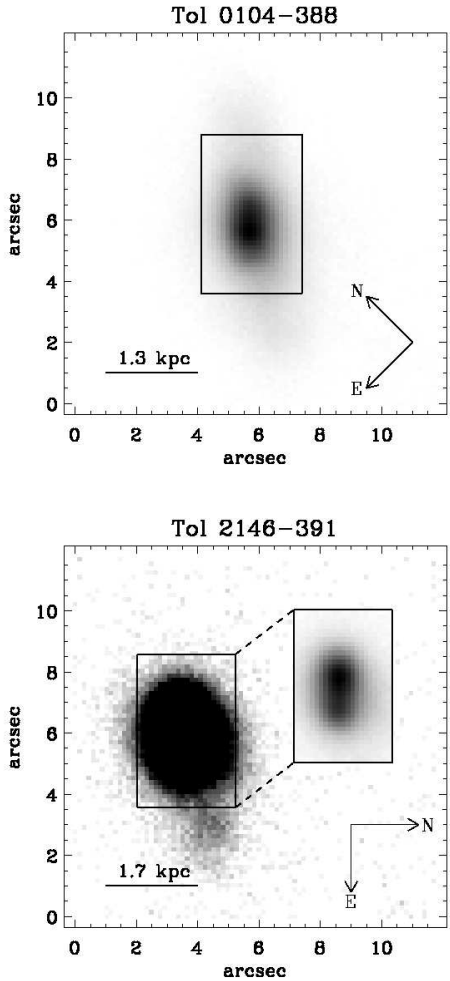}
\caption{g-band acquisition image of the galaxies Tol 0104-388 and Tol 2146-6391 in logarithmic scale.  
The rectangle indicates the GMOS--IFU FoV of 3\arcsec.5$\times$5\arcsec.
}
\label{image}
\end{figure}

To date, a small but an increasing number of H\,{\sc ii}/BCD galaxies has been 
observed using Integral Field Unit (IFU) spectroscopy \cite[e.g.,][]{I06a,L09,C09a,C09b,C10,MI10} 
in order to study the spatial distribution of their properties 
(i.e., emission lines, extinction, kinematics, abundances).
In particular, some of these studies have shown a remarkable chemical 
homogeneity of oxygen abundance in BCD galaxies \citep[e.g.,][]{L09}.
This implies that oxygen and all $\alpha$-elements are primary, produced by massive stars
($>$8-10 M$_{\odot}$) and released into the interstellar medium (ISM) during their supernova (SN) phase. 
These newly synthesized metals will be dispersed in the whole galaxy and mixed 
via hydrodynamic mechanisms in timescales of few 10$^{8}$ yr.
On the other hand, the ratio N/O was also found to be rather constant for BCD galaxies 
\citep[log(N/O)$\simeq$-1.6;][]{EP78,A79,I99}, implying mainly primary 
production of nitrogen at low metallicity (12+log(O/H)$\leq$7.6), 
although the amount coming from each source, primary or secondary, 
is still debated because of the lack of a clear mechanism that produces N in
massive stars besides the effect of the stellar rotation
\citep{MM05}.
The observed spread of N/O at 7.6$\leq$12+log(O/H)$\leq$8.3
is seen to be large and has been attributed to observational uncertainties, a loss of heavy elements 
via galactic winds \citep{V98} and/or to the time delays (delayed-release hypothesis) 
between the production of oxygen by massive stars 
and that of nitrogen by intermediate and/or massive stars. In this scenario the 
SF is an intermittent process in H\,{\sc ii}/BCD galaxies \citep{G90} with several SF bursts.
However, the delayed-release scenario cannot explain the presence of H\,{\sc ii}/BCD 
galaxies with high N/O ratio in comparison to the expected value for their O content.
The most plausible explanation of the high N/O ratio observed in these
objects is the chemical pollution of N due to the presence of Wolf-Rayet
(WR) stars \citep[e.g.,][]{LS10a}. Localized nitrogen self-enrichment has been measured in a few cases 
\citep[e.g.,][in NGC 5253]{K97} and attributed to the release of N into the ISM by the action 
of strong winds produced by these stars. Although, there are a few BCD galaxies with
significant variations of oxygen abundance, such as in two HII regions of
Haro 11 \citep[12+logO/H=8.33$\pm$0.01 in Haro B and 12+logO/H=8.10$\pm$0.04 in Haro C;][]{G12},
no localized oxygen enrichment systems have been confirmed \citep[][]{L09}.
Finally, at high metallicity (12+log(O/H)$\geq$8.3) 
the N/O ratio clearly increases with increasing oxygen abundance. Hence, nitrogen
is essentially a secondary element in this metallicity regime.

The origin of hard-ionization emission lines such as He \,{\sc ii} $\lambda$4686 
in BCD galaxies has been a subject of study in recent years given that photoionized models of H\,{\sc ii} regions 
fail to reproduce the observed high-ionization line ratios.
In particular, the observed intensity of He \,{\sc ii} $\lambda$4686 with respect to H$\beta$ is several orders of 
magnitude larger than model predictions for photoionized regions \citep{S90}.
Several mechanisms for producing hard ionizing radiation have been proposed, 
such as massive main-sequence stars \citep{S97}, WR stars \citep{S96}, primordial zero-metallicity stars 
\citep[][]{S02,S03}, high-mass X-ray binaries \citep[HMXBs;][]{G91}, radiative shocks \citep{DS96} and O stars 
at low metallicity may also contribute to the He \,{\sc ii} ionizing flux \citep{B08}. 
\citet{F01} and \citet[][]{I01,I04} have explored these different mechanisms 
which can produce the hard radiation in SBS 0335-052, Tol 1214-277 and Tol 65. 
They concluded that the ionization produced by main-sequence stars cannot explain 
the strong [Fe V] $\lambda $4227 and He \,{\sc ii} $\lambda $4686 emission lines, but other ionization 
sources as WR stars, HMXB systems and fast shocks, can be considered.

We focus here on two compact H\,{\sc ii} galaxies: Tol 0104-388 and Tol 2146-391 in order 
to study the relationship between the physical-chemical and kinematics properties, 
the nature of their hard ionization radiation pattern 
(He \,{\sc ii} $\lambda$4686 emission lines) in the ISM of these galaxies based 
on IFU spectroscopic observations on the Gemini South telescope. 
We selected these galaxies because both objects belong to a subset sample of H\,{\sc ii} galaxies with 
compact morphology and relative low metallicity (12+log(O/H)$\la$7.8-7.9 dex), possibly mimicking the properties 
one expects for young galaxies at high redshift. 
In Fig.~1 we show g-band acquisition images of the galaxies Tol 0104-388 and Tol 2146-391. 
With the present observations, we note that Tol 2146-391 encompasses two main giant H\,{\sc ii} regions (GH\,{\sc ii}Rs), as shown 
in Fig.~1. While Tol 0104-388 shows one GH\,{\sc ii}R.
In Table ~1 we show the general parameters of these galaxies.

\begin{table*}
 \centering
 \begin{minipage}{140mm}
  \caption{General parameters of the galaxies.}
  \begin{tabular}{@{}lccccccrlr@{}}
  \hline
   Name     &    \multicolumn{2}{l}{Coordinates (J2000)}& D (3K CMB)\footnote{Obtained from NED.} 
&  1\arcsec (3K CMB) & 12+log(O/H)\footnote{Derived from the present observations.} & Other names \\       
   
 \hline
Tol 0104-388 & 01:07:02.1 &  -38:31:52 & 88.4 & 429 &8.02 & CTS 1001 \\ 
Tol 2146-391 & 21:49:48.2 &  -38:54:09 & 117.3& 569 &7.82 & \\

\hline
\end{tabular}
\end{minipage}
\end{table*}

This paper is organized as follows: the observations and data reduction
are presented in Sect.~2. In Sect.~3 we show our results based in the study 
of the detected emission lines. In Sect.~4 we discuss our results and in 
Sect.~5 we summarize our conclusions.

\section{Observations and Data reduction}\label{observation}

The observations were performed with the Gemini Multi-Object Spectrograph GMOS
\citep{Hook04} and the IFU unit \citep[][hereafter GMOS--IFU]{Allington02}
at Gemini South Telescope in Chile, using the grating B600$+\_$G5323 (B600) 
and grating R600$+\_$G5324 (R600) in one slit mode. 
The GMOS-IFU in one slit mode composes a pattern
of 750 hexagonal elements with a projected diameter of 0\arcsec.2, covering a total
3\arcsec.5 $\times$5$\arcsec$ field of view (FoV), where 250 of these elements 
are dedicated to sky observation. The CCD detector is assembled by 3 chips with two small gaps.
The data were obtained in two different nights 
with seeing that range values from $\sim$0$\arcsec$.7 to $\sim$0$\arcsec$.9
obtained from the FWHM of  stars in acquisition images.
Table ~2 summarizes our observations.

\begin{table*}
 \centering
 \begin{minipage}{100mm}
  \caption{Observing log.}
  \begin{tabular}{@{}lccccclrlr@{}}
  \hline      
 Name   &    Grating & Date &Exp. time & Airmass\footnote{Obtained from the average values of the diferent exposure.} & 
Seeing\footnote{The seeing was obtained as the FWHM of  stars in acquisition images.}\\
        &            &      & (s)      &         &  (\arcsec)     \\
 \hline
Tol 0104-388 & B600&2004-12-31&3$\times$1800&1.45 &0.69\\
             & R600&2005-10-07&3$\times$2400&1.03 &0.86\\ 
Tol 2146-391 & B600&2005-08-11&3$\times$1800&1.19 &0.91\\
             & R600&2004-09-20&2$\times$2400&1.02 &0.82\\
\hline
\end{tabular}
\end{minipage}
\end{table*}

Data reduction was carried out using the Gemini software package version 1.9 inside IRAF\footnote{IRAF
is distributed  by NOAO, which is operated by the Association of
Universities for Research in Astronomy Inc., under cooperative agreement
with the National Science Foundation.}. This includes bias subtraction, 
flat-field correction, wavelength calibration and sky subtraction. 
The flux calibration was performed using the sensitive function derived from
observation of the stars EG 21 and H 600.
The 2D data images were transformed into 3D data cubes, resampled as square pixels with 
0\arcsec.2 spatial resolution and corrected for differential atmospheric refraction (DAR) 
using the \textit{gfcube} routine. Finally, the data cubes obtained using the gratings 
B600 and R600 were combined, for each galaxy, forming a final data cube covering a total spectral 
range from $\sim$3021 to 7225 $\rm\AA$.
More details about the reduction procedure applied to our data can be found in \cite{L09}.

\begin{figure*}
\includegraphics[width=175mm]{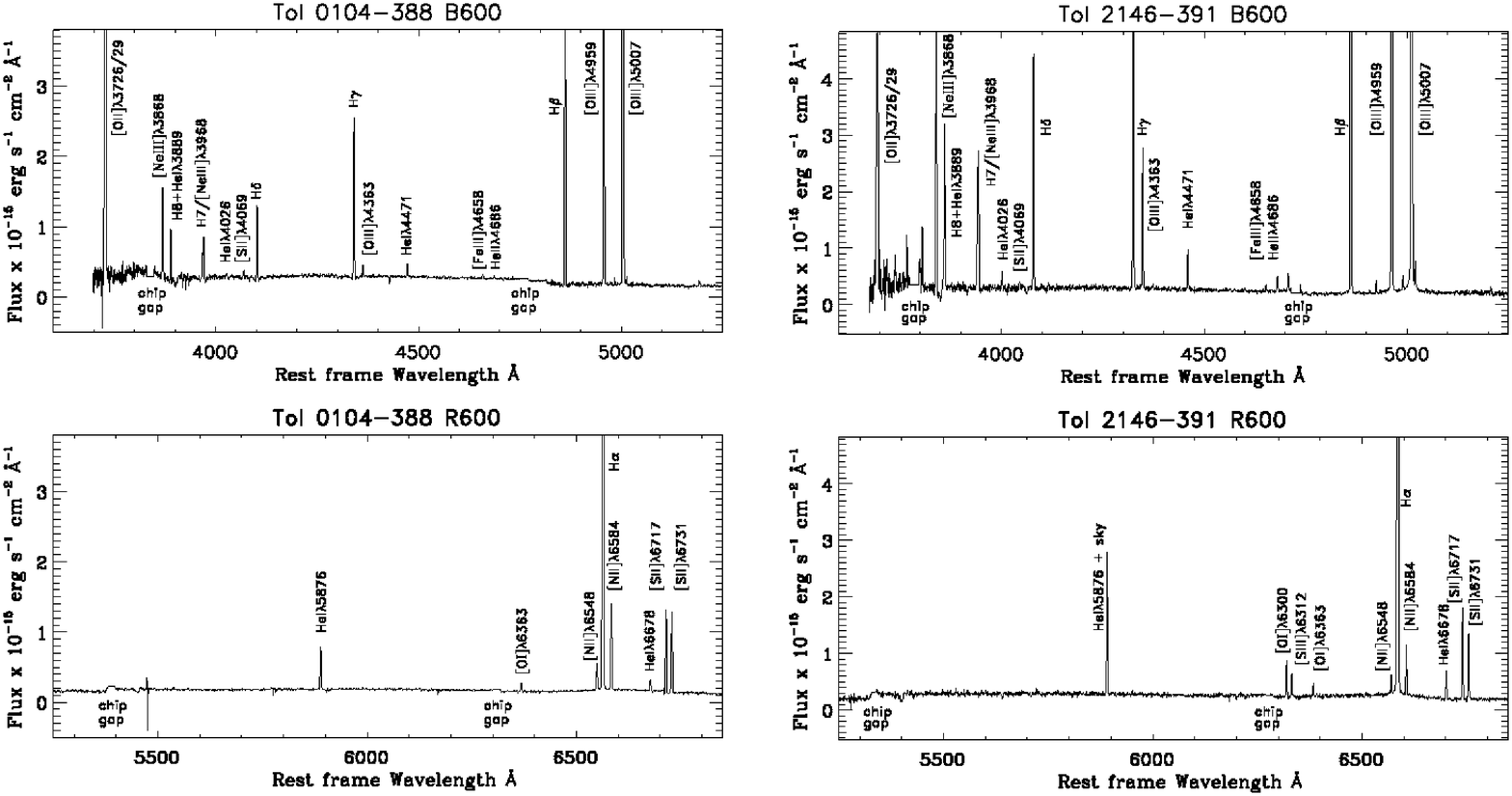}
\caption{Integrated spectra (summing over all 
spaxels in the FoV) of Tol 0104-388 and Tol 2146-391. Top panels: blue spectrum (B600 grating), 
Bottom panels: red spectrum (R600 grating). The gaps between the detectors are shown in the figure.
}
\label{integrated_spec}
\end{figure*}

Figure \ref{integrated_spec} shows the integrated spectrum of the galaxies summing over all 
spaxels in the FoV.
In this figure we identified the main emission lines detected and used in our study.
The spectral resolution (R$_{B600}$=1688 and R$_{R600}$=3744) of the GMOS--IFU using gratings B600 and R600 allows us to resolve 
the [O \,{\sc ii}] $\lambda\lambda$3726/29 doublet, [O \,{\sc iii}] $\lambda$4363 and other weak emission lines. 
The most remarkable feature in the spectrum of these galaxies
is the detection of He \,{\sc ii} $\lambda$4686 that are associated with a high-ionization radiation. 
We detect, for the first time, very weak He \,{\sc ii} $\lambda$4686 emission line in the spectrum of Tol 0104-388. 
While in Tol 2146-391 the presence of this high ionization emission
line has been measured previously in the literature \citep[e.g.,][]{P06,G07,G11}.   

Finally, the emission line fluxes were measured using the IRAF task \textit{fitprofs} 
by fitting Gaussian profiles. 
Since most of the emission lines were measured using an automatic procedure, we filtered the maps 
assigning the value 0 erg cm$^{-2}$ s$^{-1}$ to all spaxels with signal to noise ratio (S/N) $<$ 3.
The error associated with each emission line ($\sigma_{i}$) 
was calculated using that $\sigma _{i}=\sigma _{c} N^{1/2} \sqrt{1+EW/N\Delta}$,
where $\sigma_{c}$ is the standard deviation in the local continuum associated with each emission line, 
N is the number of pixels, EW is the equivalent width of the line, and $\Delta$ is the instrumental 
dispersion in $\rm \AA$ \citep[see][]{L09}.

\section{Properties of the ionized gas}\label{results}

\subsection{Emission lines, EW(H$\beta$) and extinction} \label{emission}

\begin{figure*}
\includegraphics[width=175mm]{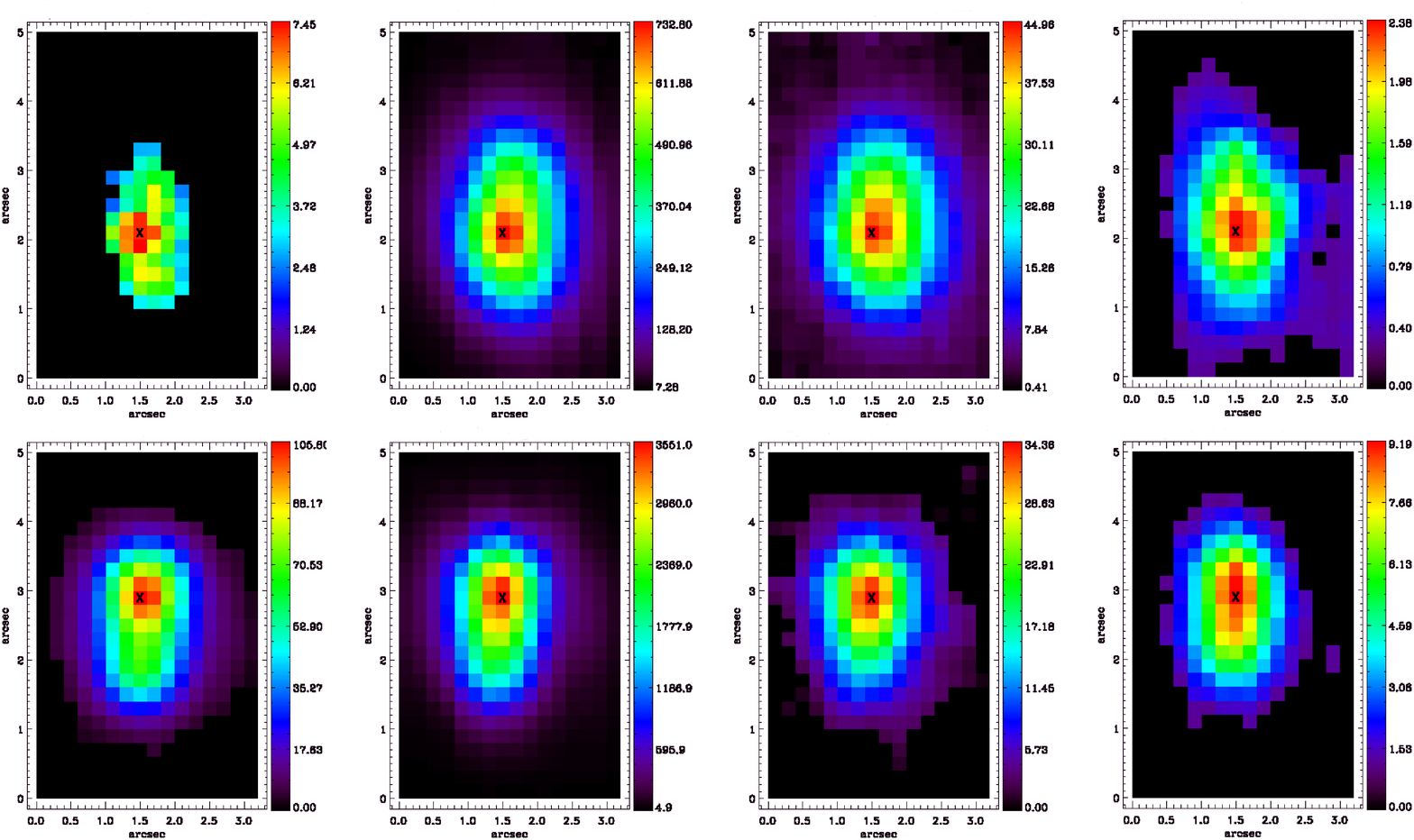}
\caption{
Observed emission line maps (from the left to the right) [O\,{\sc iii}]$\lambda$4363, H$\alpha$,
[N\,{\sc ii}]$\lambda$6584, and
H$\alpha$ continnum of Tol 0104-388 (upper panels; 1\arcsec=429 pc) and Tol 2146-391 (lower panels; 1\arcsec=569 pc) 
in the original 0\arcsec.2 pixel size. Fluxes in units of  10$^{-18}$ ergs cm$^{-2}$ s$^{-1}$ and
continuum in units of 10$^{-18}$ ergs cm$^{-2}$ s$^{-1}$ $\rm \AA^{-1}$. 
The maximum H$\alpha$ emission is indicated in the maps by an X symbol.
}
\label{mapas}
\end{figure*}

We used the flux measurement procedure described previously in Sect.~2 
to construct emission line maps (e.g., [S \,{\sc ii}] $\lambda$6731, [S \,{\sc ii}] $\lambda$6717, 
[N \,{\sc ii}] $\lambda$6548, [N \,{\sc ii}] $\lambda$6584, H$\alpha$, [O \,{\sc iii}] $\lambda$5007, 
[O \,{\sc iii}] $\lambda$4959, H$\beta$, He \,{\sc ii} $\lambda$4684, [O \,{\sc iii}] $\lambda$4363, 
H$\gamma$, [O \,{\sc ii}] $\lambda$3726,29, continuum) in a FoV of 3\arcsec.2$\times$5\arcsec, 
equivalent to $\sim$1372 pc $\times$ 2058 pc for Tol 0104-388 and 
$\sim$1820 pc $\times$ 2730 pc for Tol2146-391. 
In Fig. \ref{mapas} we show the [O \,{\sc iii}] $\lambda$4363, H$\alpha$, 
[N \,{\sc ii}] $\lambda$6586 emission line maps and continuum near H$\alpha$ of  
Tol 0104-388 (upper maps) and Tol2146-391 (lower maps), respectively. 
The flux map of these emission lines displays a single peak in Tol 0104-388 but 
the elongated shape of Tol 2146-391 suggests a double peak as is noted in the acquisition image 
of Fig. \ref{image}. Unfortunately, we did not resolve 
spatially these two GH\,{\sc ii}Rs in our monochromatic maps from the data cube of Tol 2146-391.

The observed emission line fluxes, for all the spaxels in our FoV, 
have been corrected for extinction using the observed Balmer
decrement as $I(\lambda)/I(H\beta)=F(\lambda)/F(H\beta) \times 10^{c(H\beta)f(\lambda)}$,
using the \cite{C89} reddening curve, f($\lambda$). 
The observed F($\lambda$) and corrected emission line fluxes I($\lambda$) 
relative to the H$\beta$ multiplied by a factor of 100 and their error, the observed flux of the H$\beta$ emission line, 
the H$\beta$ equivalent widths EW(H$\beta$) and the extinction coefficient c(H$\beta$) for the integrated galaxy 
are given in Table 3.

\begin{table*}
 \centering
 \begin{minipage}{100mm}
  \caption{Observed and extinction corrected emission lines for Tol 0104-388 and Tol 2146-391. 
The fluxes are relative to F(H$\beta$)=100.}
  \begin{tabular}{@{}lcccc@{}}
  \hline
        &   \multicolumn{2}{c}{Tol 0104-388} &   \multicolumn{2}{c}{Tol 2146-391}\\
 & F($\lambda$)/F(H$\beta$) & I($\lambda$)/I(H$\beta$) & F($\lambda$)/F(H$\beta$) & I($\lambda$)/I(H$\beta$)\\
 \hline
$\left[O II\right] \lambda$3726    & 76.00$\pm$2.08 & 96.37$\pm$5.29 & 18.84$\pm$2.31 & 23.54$\pm$2.76 \\
$\left[O II\right] \lambda$3729    &105.30$\pm$3.30 &133.47$\pm$8.37 & 25.72$\pm$0.23 & 32.12$\pm$0.27 \\
$\left[Ne III\right] \lambda$3868  & 23.80$\pm$0.99 & 29.50$\pm$2.45 & 28.98$\pm$1.28 & 35.44$\pm$1.50 \\
H8+He I $\lambda$3889              & 13.20$\pm$0.51 &16.30$\pm$1.27 & 13.00$\pm$0.40 & 15.84$\pm$0.97    \\ 
$\left[Ne III\right] \lambda$3968  &  6.80$\pm$0.53 &  8.27$\pm$1.28 &  9.46$\pm$0.16 & 11.51$\pm$0.39 \\
H7  $\lambda$3970                  &  9.26$\pm$0.50 & 11.26$\pm$1.23 &  9.76$\pm$0.13 & 11.88$\pm$0.32 \\
He I $\lambda$4026                 &  1.37$\pm$0.30 &  1.65$\pm$0.72 &  1.18$\pm$0.07 &  1.40$\pm$0.17  \\ 
$\left[S II\right] \lambda$4069    &  2.09$\pm$0.25 &  2.49$\pm$0.59 &  0.85$\pm$0.58 &  1.00$\pm$0.66 \\
H$\delta  \lambda$4101             & 19.80$\pm$2.90 & 23.45$\pm$6.88 & 19.02$\pm$1.04 & 22.29$\pm$1.18 \\
H$\gamma  \lambda$4340             & 42.60$\pm$2.13 & 47.81$\pm$4.78 & 40.76$\pm$2.07 & 45.42$\pm$2.26 \\
$\left[O III\right] \lambda$4363   & 3.60$\pm$0.20 & 4.02$\pm$0.45  & 10.78$\pm$0.63 & 11.95$\pm$0.68\\
He I $\lambda$4471                 & 3.63$\pm$0.11 &  3.95$\pm$0.25 &  3.22$\pm$0.09 & 3.49$\pm$0.19\\ 
$\left[Fe III\right] \lambda$4658  & 1.38$\pm$0.05 &  1.44$\pm$0.11  &  0.67$\pm$0.04 & 0.70$\pm$0.08\\
He II $\lambda$4686                & 0.80$\pm$0.09 &  0.83$\pm$0.18 &  1.56$\pm$0.52 &  1.61$\pm$0.53\\
H$\beta \lambda$4861               &100.00$\pm$3.70          & 100.00$\pm$3.70         &100.00$\pm$2.38 &100.00$\pm$2.38\\
$\left[O III\right] \lambda$4959   &119.33$\pm$5.23          & 117.09$\pm$5.13         &202.90$\pm$4.09 &199.33$\pm$4.03 \\
$\left[O III\right] \lambda$5007   &353.33$\pm$17.69         & 343.64$\pm$17.20        &610.51$\pm$10.47&594.79$\pm$10.25\\
He I $\lambda$5876                 & 12.27$\pm$0.36          &  10.56$\pm$0.62         &   $\cdots$     &  $\cdots$       \\ 
$\left[O I\right] \lambda$6300     &$\cdots$                 &    $\cdots$             &  2.83$\pm$0.40 &  2.36$\pm$0.67\\
$\left[S III\right] \lambda$6312    &$\cdots$                 &    $\cdots$             &  1.62$\pm$0.06 &  1.35$\pm$0.10\\
$\left[O I\right] \lambda$6363     & 2.72 $\pm$0.24          &   2.23$\pm$0.39         &  0.95$\pm$0.03 &  0.79$\pm$0.05\\
$\left[N II\right] \lambda$6548    &  7.93$\pm$1.05          &   6.38$\pm$0.84         &  1.90$\pm$0.25 &  1.55$\pm$0.21 \\
H$\alpha   \lambda$6563            &370.00$\pm$10.06         & 297.15$\pm$8.08         &362.32$\pm$7.22 &295.00$\pm$6.12 \\
$\left[N II\right] \lambda$6584    & 25.20$\pm$1.60          &  20.20$\pm$1.28         &  3.86$\pm$0.42 &  3.14$\pm$0.35 \\
He I $\lambda$6678                 &  3.59 $\pm$0.16         &   2.85$\pm$0.25         &  2.06$\pm$0.06 &  1.66$\pm$0.10\\ 
$\left[S II\right] \lambda$6717    & 25.20$\pm$1.68          &  19.94$\pm$1.33         &  7.06$\pm$0.91 &  5.67$\pm$0.76\\
$\left[S II\right] \lambda$6731    & 24.86$\pm$1.64          &  19.64$\pm$1.29         &  5.22$\pm$0.81 &  4.18$\pm$0.68 \\
                                   &       \\
F(H$\beta$)\footnote{In units of $\times$10$^{-15}$ erg cm$^{-2}$ s$^{-1}$}& 15.00$\pm$0.28  & & 55.20$\pm$0.61\\
EW(H$\beta$)\footnote{In units of $\rm \AA$}                               & 109$\pm$22  & & 229$\pm$30       \\
c(H$\beta$)                       & 0.32 & & 0.30  \\
\hline
\end{tabular}
\label{tb_lines_1}
\end{minipage}
\end{table*}

\begin{figure}
\includegraphics[width=85mm]{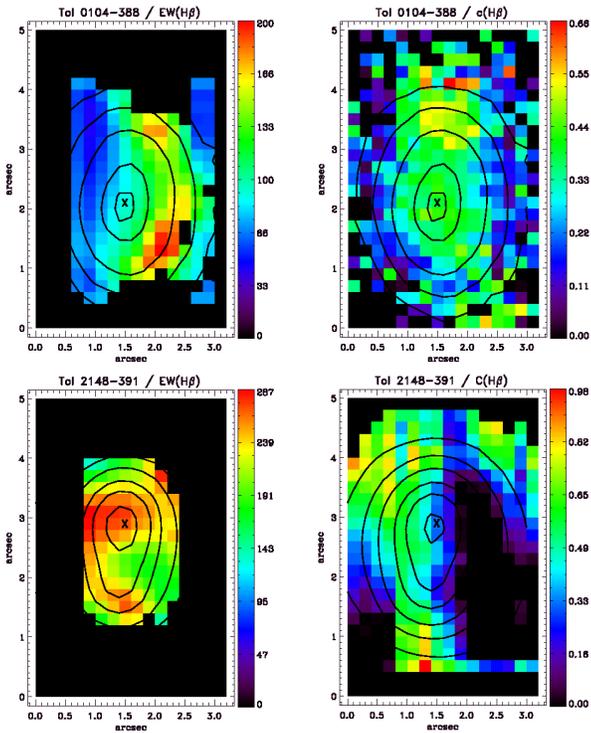}
 \caption{Spatial distribution of EW(H$\beta$) and c(H$\beta$) for our sample galaxies.
The maximum H$\alpha$ emission is indicated in the maps by an X symbol. H$\alpha$ emission line 
contours are overplotted on all maps.}
 \label{EW_C}
\end{figure}

In Fig. \ref{EW_C} we show the spatial distribution of EW(H$\beta$) and extinction c(H$\beta$) 
for our sample galaxies.
The EW(H$\beta$) in Tol 0104-388 
varies from $\sim$0 $\rm \AA$ to 200 $\rm \AA$ 
and in Tol 2146-391 varies from $\sim$0 $\rm \AA$ to 386 $\rm \AA$. 
While, the c(H$\beta$) extinction varies from 0 to $\sim$0.66 in Tol 0104-388 
and from 0 to $\sim$1.33 in Tol 2146-391. 
The EW(H$\beta$) and extinction distribution c(H$\beta$) in these galaxies shows an inhomogeneous pattern, 
where the peak of these values are 
displaced from the maximum of H$\alpha$ emission. In particular, the extinction map of Tol 2146-391 shows
a hole-like structure in the right part of the FoV, where several spaxels located in this regions shows values close to 0 
or F(H$\alpha$)/F(H$\beta$)$\sim$2.87. 
These structures appear to be real given that are not produced by misalignment between the maps.

We compared our integrated extinction value of c(H$\beta$)=0.32, in Tol 0104-388, that is in reasonable 
agreement with those derived with long slit spectroscopy of 0.30 by \cite{P91} and 0.39 by \cite{T97}.
The integrated He \,{\sc ii} $\lambda$4686 emission in Tol 0104-388 is very weak and  corresponds to $\la$1\% 
of the observed H$\beta$ flux. 
In the case of Tol 2146-391, we obtained an integrated 
extinction c(H$\beta$)= 0.30 that differs slightly from the value obtained by \citet{G07} of 0.18, 0.20 by \cite{P06} 
and 0.12 by \citet{K06}, but agrees with the value found by \cite{G11} of 0.30 for the eastern 
and 0.32 for the western H\,{\sc ii} regions, respectively. 
The integrated He \,{\sc ii} $\lambda$4686 emission in Tol 2146-391 corresponds 
to $<$2\% of the observed H$\beta$ flux. Our intensity value of 
I(He \,{\sc ii} $\lambda$4686)/I(H$\beta$)=0.016$\pm$0.005, in this galaxy, agrees within the errors 
with the value of I(He \,{\sc ii} $\lambda$4686)/I(H$\beta$)=0.017$\pm$0.010 obtained by \citet{P06}, 0.022$\pm$0.001
by \cite{G07} and 0.017$\pm$0.001 for the western and 0.023$\pm$0.001 for the eastern GH\,{\sc ii}Rs by \cite{G11}. 

Finally, we estimated the mean age of the starburst in each one of the galaxies comparing 
the integrated EW(H$\beta$)=109 $\rm \AA$ for Tol 0104-388 and EW(H$\beta$)=229 $\rm \AA$ 
for Tol 2146-391 with STARBURST99 \citep{L99}. We assumed an instantaneous burst, a metallicity 
of Z=0.004 and a Kroupa IMF ($\propto$M$^{-\alpha}$) with $\alpha$=1.3 
for stellar masses between 0.1 to 0.5M$_{\odot}$ and $\alpha$=2.3 for masses 
between 0.5 and 100M$_{\odot}$. Thus, the estimated mean age in the FoV of the galaxies 
is $\sim$5 Myr in Tol 0104-388 and $\sim$4 Myr in Tol 2146-391. 

\subsection{Velocity field}\label{velocity_field}

\begin{figure}
\includegraphics[width=84mm]{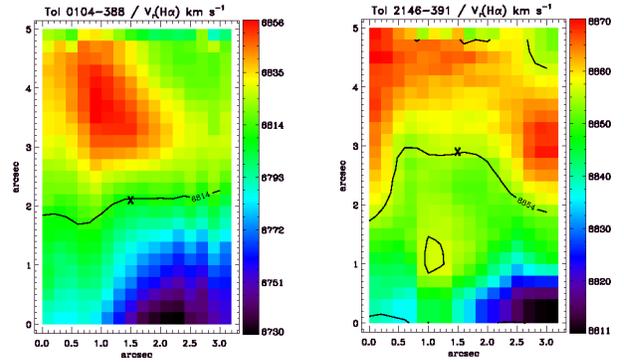}
 \caption{H$\alpha$ radial velocity in units of km s$^{-1}$.
The maximum H$\alpha$ emission is indicated in the maps by an X symbol. 
The black contours in the radial velocity maps
correspond to the systemic velocity.}
\label{velocidad}
\end{figure}

We have obtained the spatial distribution of radial velocity v$_{r}$
by fitting a single Gaussian to the H$\alpha$ emission line profiles. 
In order to better examine the variations in the FoV, in Fig.~\ref{velocidad} we show 
the smoothed radial velocity derived from the shifts of the H$\alpha$ line peak.

The velocity field v$_{r}$ of Tol 0104-388 in the left panel of Fig. \ref{velocidad} 
shows an apparent systemic motion where the upper part of the galaxy is redshifted, while the
lower part is blueshifted, with a relative motion of $\sim$50 km s$^{-1}$ 
with respect to the systemic velocity (black contour in the radial velocity map). 
This velocity map indicates the presence of a clear rotation pattern in this galaxy.
On the other hand, the radial velocity map of Tol 2146-391 (right panel of Fig. \ref{velocidad}) is rather complex 
indicating asymmetric gas motions without any clear rotational pattern. 
The whole range of radial velocities displayed in the map is about 60 km s$^{-1}$. 
No significant differences were found from the map derived using H$\beta$ and [O\,{\sc iii}]$\lambda$5007 in both galaxies. 

The systemic velocity obtained from the fit to the emission maxima (located on the cross in the velocity maps) 
is 6803  km s$^{-1}$ for Tol 0104-388 and 8854  km s$^{-1}$ for Tol 2146-391. 
For the integrated spectrum of the galaxies we obtained values of 6815 km s$^{-1}$
for Tol 0104-388 and 8854 km s$^{-1}$ for Tol 2146-391, respectively. 
Although our GMOS--IFU observations were performed using the medium resolution grating, 
we found a velocity dispersion\footnote{We derived 
the velocity dispersion using the relationship 
$\sigma^{2}$=$\sigma_{obs}^{2}$-$\sigma_{inst}^{2}$-$\sigma_{th}^{2}$. 
We used a value of 33.4 $\pm$ 1.0 km s$^{-1}$ for the instrumental dispersion 
and $\sigma_{th}$=$\sqrt{kT_{e}/m_{H}}\approx$10 km s$^{-1}$ and 11 km s$^{-1}$ for thermal broadening
considering T$_{e}$=12199 K and 15277 K  for Tol 0104-388 and Tol 2146-391, respectively.}
of $\sigma$(H$\alpha$)$\simeq$ 44  km s$^{-1}$ in Tol 0104-388 
and $\sim$25 km s$^{-1}$ in Tol 2146-391 assuming a simple Gaussian profile.  
These results are in agreement with the integrated values obtained by \citet{B11} 
with $\sigma$(H$\alpha$)=48.2$\pm$0.4 km s$^{-1}$ in Tol 0104-388 
and $\sigma$(H$\alpha$)=25.6$\pm$0.4 km s$^{-1}$ in Tol 2146-391,
using the high dispersion spectrograph FEROS at ESO. 

\subsection{Emission Line Diagnostic Diagrams}\label{diagramas}

\begin{figure}
\centering
\includegraphics[width=80mm]{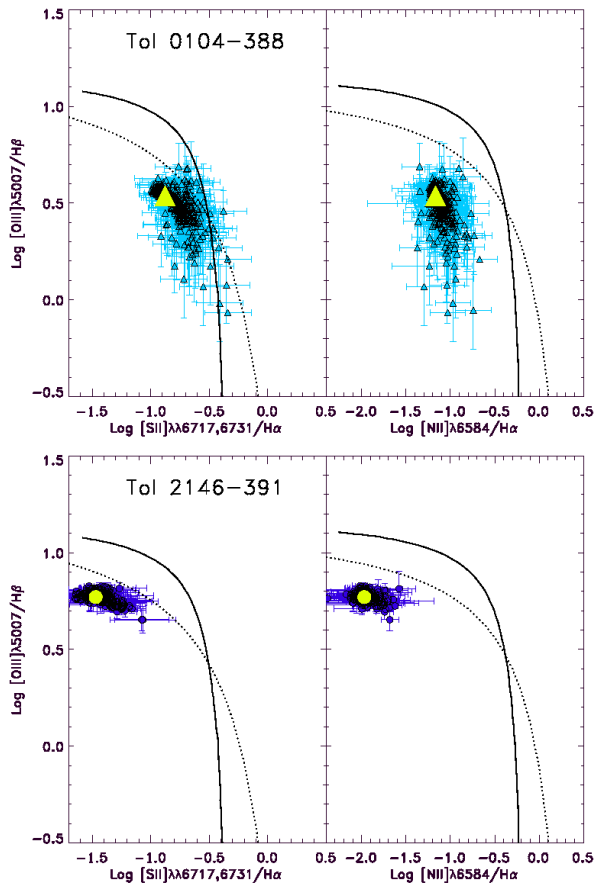}
 \caption{Position of the individual 0\arcsec.2 spaxels in Tol 0104-388 (upper panels) and 
Tol 2146-391 (lower panels) in the BPT diagnostic diagrams.
The solid curves show the empirical borders found by \citet{O06}, while the dotted lines show 
the theoretical borders proposed by \citet{K01}. The grey filled circles and triangless
mark the integrated values of the galaxies.}
\label{emi_ratios}
\end{figure}

The standard diagnostic diagrams BPT \citep{B81} have been used to analyze 
the possible excitation mechanisms present in Tol 0104-388 and Tol 2146-391. 
Fig. \ref{emi_ratios} considers: [O III]$\lambda$5007/H$\beta$ versus 
[S II]$\lambda\lambda$6717,6731/H$\alpha$ and  [O III]$\lambda$5007/H$\beta$ versus [N II]$\lambda$6584/H$\alpha$
for both galaxies in our sample.
The solid lines \citep[adapted from][]{O06} show the locus of separation between 
regions dominated solely by photoionization (HII-like regions) and regions dominated by shocks (e.g., AGNs). 
The dotted line represents the same as the solid line but using the models given by \citet{K01}. 
This fig. shows that the position of all individual spaxels, in these diagrams, suggest that photoionization 
from stellar sources is the dominant excitation mechanism in our two galaxies. 

\begin{figure*}
\includegraphics[width=140mm]{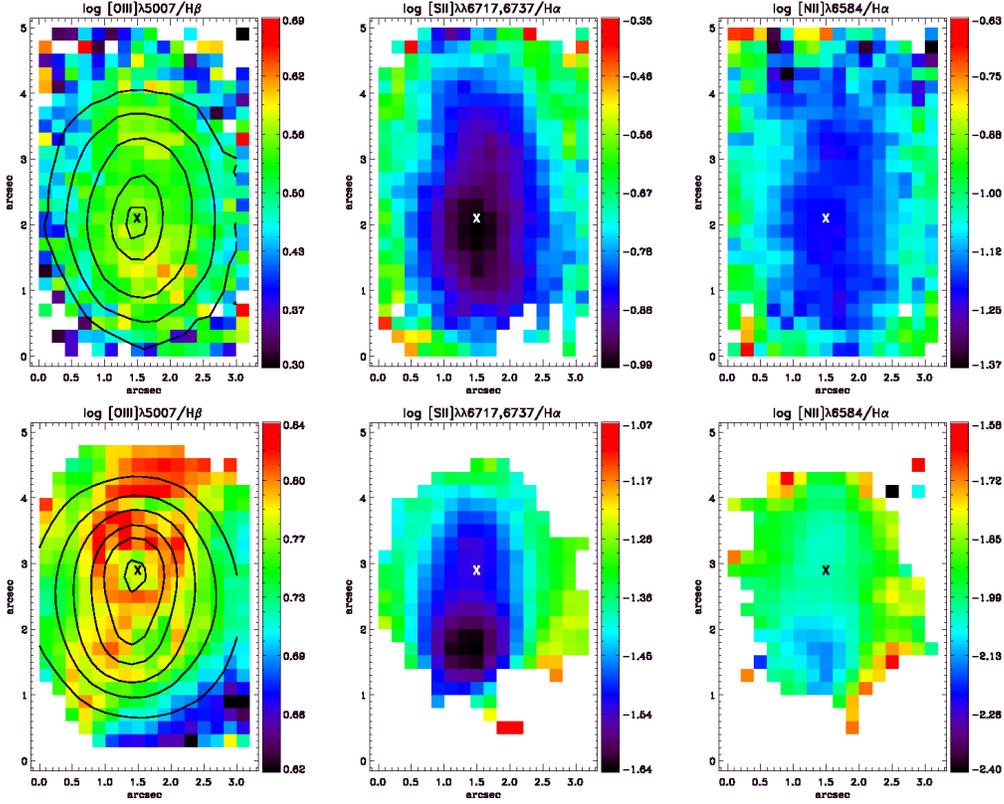}
 \caption{Emission line ratios: log [O III]$\lambda$5007/H$\beta$, 
log [S II]$\lambda\lambda$6717,6731/H$\alpha$ and log [N II]$\lambda$6584/H$\alpha$ 
for both galaxies in our sample (Tol 0104-388: upper pannels and Tol 2146-391: lower pannels). 
The maximum H$\alpha$ emission is placed over region A and is indicated in the maps by an X symbol.
}
\label{emi_ratios_spa}
\end{figure*}

In Fig. \ref{emi_ratios_spa} we show the maps for the three available line ratios involved in these diagrams 
[O III]$\lambda $5007/H$\beta $, [S II]$\lambda\lambda$6717,6731/H$\alpha$ and [N II]$\lambda$6584/H$\alpha$. 
This fig. shows that the ionization structure in the central region of Tol 0104-388 
is rather constant for [N II]$\lambda$6584/H$\alpha$. 
While [O III]$\lambda $5007/H$\beta$ and [S II]$\lambda\lambda$6717,6731/H$\alpha$
do not show a uniform distribution. The same behavior is observed in Tol 2146-391.
The spatial structure changes from the peak of H$\alpha$ emission to the outer part of the galaxies. 
Comparing the two galaxies, we can see that the data points 
of Tol 2146-391, in Fig \ref{emi_ratios}, are more highly clustered to the upper left of the 
diagram than those of Tol 0104-388. These features usually indicate a harder ionization field either by
young star clusters or star cluster complexes at lower metallicities.
Finally, our integrated [S II]$\lambda\lambda$6717,6731/H$\alpha$ and 
[N II]$\lambda$6584/H$\alpha$ values are consistent with the ones reported in the literature by \citet{G11} in Tol 2146-391 
and inferred using the emission lines reported by \citet{K06} in Tol 0104-388.

\subsection{Chemical abundances} \label{chemical}

The first step in the abundance derivation is the calculation of the electron
temperature and density. We calculated the electron temperature T$_{e}$(O \,{\sc iii})
from the ratio [O \,{\sc iii}]$\lambda\lambda$4959,5007/[O \,{\sc iii}]$\lambda$4363 
and the electron density from [S \,{\sc ii}]$\lambda$6717/[S \,{\sc ii}]$\lambda$6731 
ratio using the five level atomic model FIVEL \citep{D87}, implemented 
under the IRAF STS package $\it{nebular}$. 

\begin{figure}
\centering
\includegraphics[width=84mm]{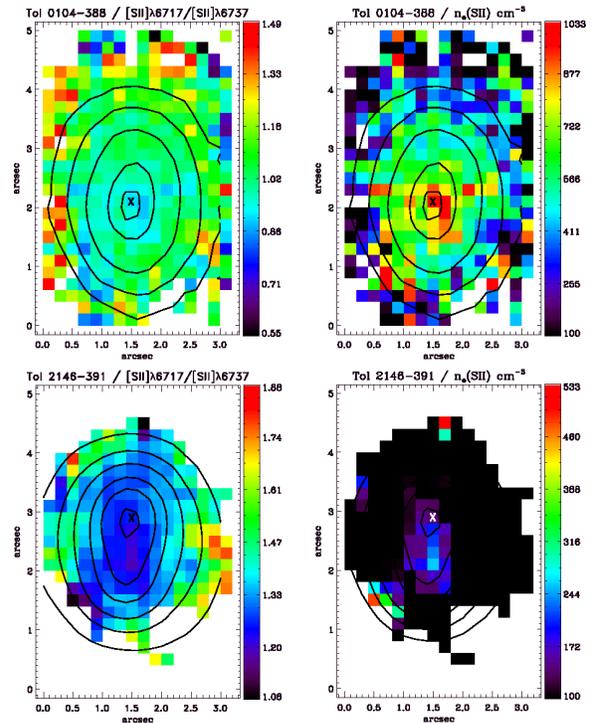}
 \caption{[S II]$\lambda$6717/[S II]$\lambda$6731 ratio maps and electron density maps 
n$_e$(S II) cm$^{-3}$ of Tol 0104-388 (upper pannels) and Tol 2146-391 (lower pannels). 
The maximum H$\alpha$ emission is indicated in the maps a X symbol. H$\alpha$ contours are overplotted on the maps.
}
\label{ratios_den}
\end{figure}

\begin{figure}
\includegraphics[width=84mm]{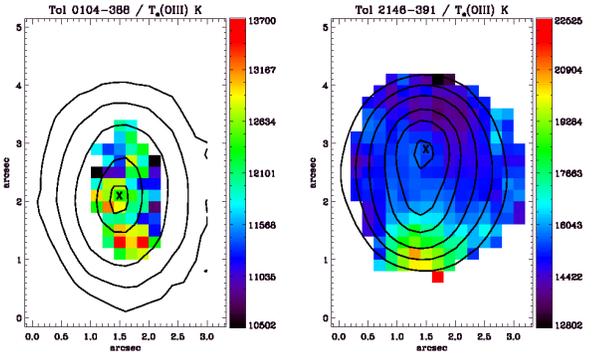}
 \caption{
Electron temperature maps in K of Tol 0104-388 (upper panel) and Tol 2146-391 (lower pannel)
obtained from [O III]$\lambda\lambda$4959,5007/[O III]$\lambda$4363 emission line ratio.
The maximum H$\alpha$ is indicated in the maps by a X symbol. 
H$\alpha$ contours are overplotted on the maps.
}
\label{temp_maps}
\end{figure}

Fig. \ref{ratios_den} shows the spatial distribution of the 
[S \,{\sc ii}]$\lambda$6717/[S \,{\sc ii}]$\lambda$6731 ratio 
and the electron density, and Fig. \ref{temp_maps} shows the 
electron temperature T$_{e}$(O \,{\sc iii}) obtained from the 
[O \,{\sc iii}]$\lambda\lambda$4959,5007/[O \,{\sc iii}]$\lambda$4363 ratio,
for both galaxies in our sample. 

\begin{figure}
\centering
\includegraphics[width=84mm]{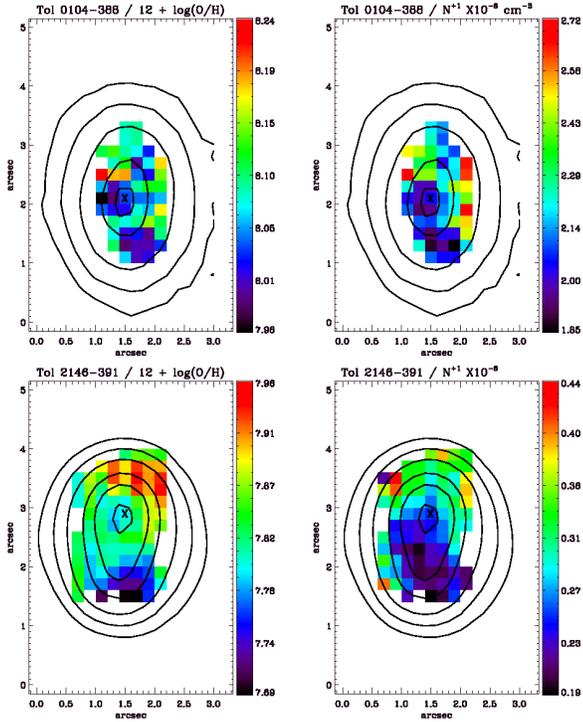}
 \caption{
12+log(O/H) abundance and N$^{+}$/H$^{+}$ ratio maps of Tol 0104-388 (upper pannels) 
and Tol 2146-391 (lower pannels). Overlaid are the H$\alpha$ flux contours. 
The maximum H$\alpha$ emission is indicated in the maps by an X symbol.
}
\label{abundance_maps}
\end{figure}

\begin{figure}
\centering
\includegraphics[width=60mm]{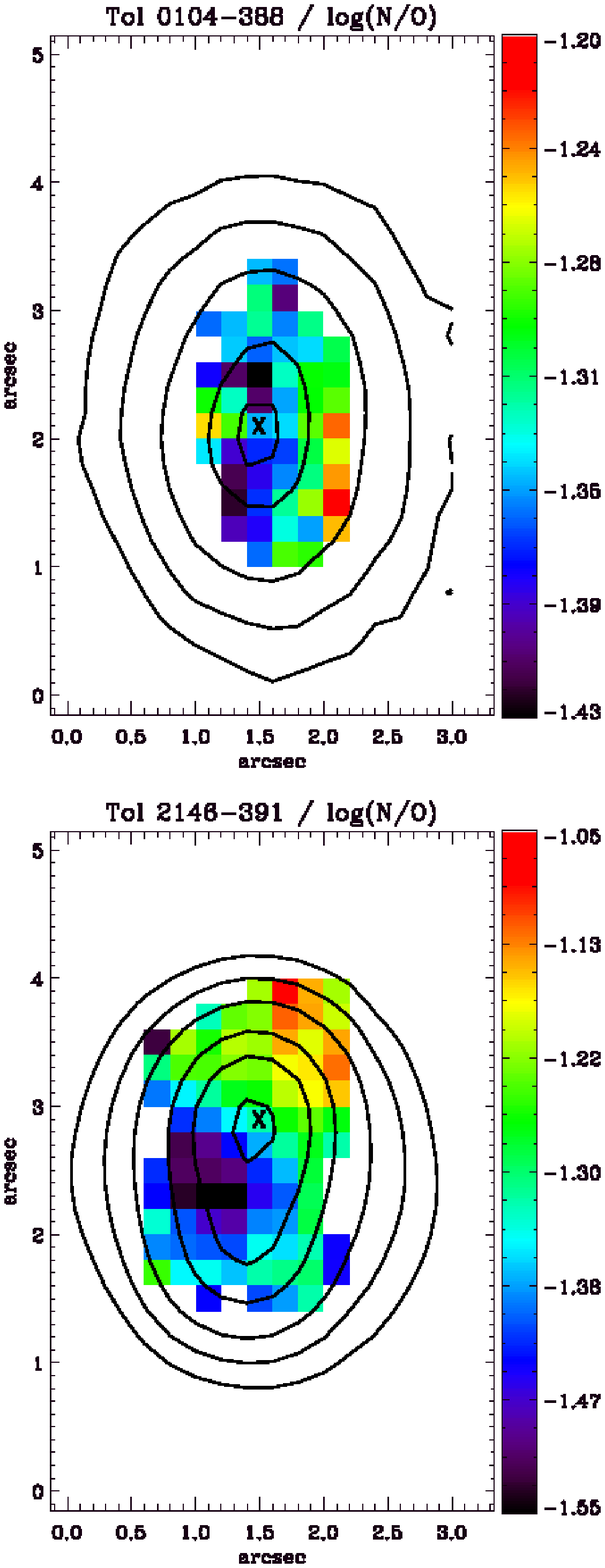}
 \caption{
log(N/O) ratio maps of Tol 0104-388 and Tol 2146-391.
Overlaid are the H$\alpha$ flux contours. 
The maximum H$\alpha$ emission is indicated in the maps by an X symbol.
}
\label{NO_maps}
\end{figure}

The oxygen, nitrogen, neon and sulfur abundances were  calculated as 

\begin{equation}
\frac{O}{H}=\frac{O^{+}}{H^{+}} + \frac{O^{+2}}{H^{+}},
\end{equation}

\begin{equation}
\frac{N}{H}=ICF(N) \frac{N^{+}}{H^{+}},
\end{equation}

\begin{equation}
\frac{Ne}{H}=ICF(Ne) \frac{Ne^{+2}}{H^{+}},
\end{equation}

\begin{equation}
\frac{S}{H}=ICF(S) \left(\frac{S^{+}}{H^{+}}+\frac{S^{+2}}{H^{+}}\right),
\end{equation}

\noindent
where  O$^{+}$, O$^{+2}$, N$^{+}$, Ne$^{+2}$ and S$^{+}$ ions are obtained from 
the \textit{nebular} output file. Nitrogen, neon and sulfur abundances were calculated using  
ionization correlation factors (ICF) and the relationship between 
S$^{+}$ and S$^{+2}$ given by \cite{KB94}. 
We assumed that T$_{e}$(O \,{\sc ii}) temperature is given by 
T$_{e}$(O \,{\sc ii}) = 2/(T$_{e}^{-1}$(O \,{\sc iii})+0.8) \citep{P92}, 
T$_{e}$(S \,{\sc iii}) = 0.83$\times$T$_{e}$(O \,{\sc iii})+0.17 \citep{G92} and 
T$_{e}$(O \,{\sc ii}) = T$_{e}$(S \,{\sc ii}) = T$_{e}$(N \,{\sc ii}). 
In Table \ref{integrated_properties_1}  
we show the electron density, temperature and abundances 
calculated for each one of the integrated apertures considered in this study. 
Fig. \ref{abundance_maps} (left panel) shows the spatial distribution 
of the oxygen abundances in the spaxels where 
the emission lines [O III]$\lambda$4363  and [O II]$\lambda\lambda$3726/29 were detected, 
and in the right panel of the same fig. we show the
spatial distribution of N$^{+}$ in the spaxels where [N II]$\lambda$6584 was detected.
Finally, in Fig. \ref{NO_maps} we show the spatial distribution of log(N/O).

\begin{table}
 \centering
 \begin{minipage}{80mm}
  \caption{Ionic, abundances and integrated properties of Tol 0104-388 and Tol 2146-391.}
  \begin{tabular}{@{}lccrrrlrlr@{}}
  \hline
                                &  Tol 0104-388 &   Tol 2146-391\\
 \hline
Te(OIII) K                      & 12199$\pm$412  &  15277$\pm$584\\
Ne(SII) cm$^{-3}$               & 614            & $\la$100\\
O$^{+}$/H$^{+} \times$10$^{5}$  &4.16$\pm$0.11   & 0.62$\pm$0.04\\
O$^{++}$/H$^{+} \times$10$^{5}$ &6.41$\pm$0.28   & 6.03$\pm$0.58\\
O/H $\times$10$^{5}$            &10.56$\pm$0.39  & 6.65$\pm$0.61\\
12+log(O/H)                     &8.02$\pm$0.04   & 7.82$\pm$0.09\\
N$^{+}$/H$^{+} \times$10$^{6}$  &2.34$\pm$0.04   & 0.32$\pm$0.01\\
ICF(N)                          &2.54$\pm$0.16   & 10.67$\pm$1.62\\
N/H $\times$10$^{6}$            &5.94$\pm$0.48   & 3.43$\pm$0.69\\
12+log(N/H)                     &6.77$\pm$0.08   & 6.54$\pm$0.10\\
log(N/O)                        &-1.25$\pm$0.12  & -1.28$\pm$0.29\\
S$^{+}$/H$^{+} \times$10$^{7}$  &2.83$\pm$0.13      & 0.97$\pm$0.05\\
S$^{++}$/H$^{+} \times$10$^{6}$ &1.67$\pm$0.10       & 0.71$\pm$0.08\\
ICF(S)                          &1.09$\pm$0.04       & 1.58$\pm$0.02\\
S/H  $\times$10$^{6}$           &2.12$\pm$0.09       & 1.28$\pm$0.02\\
12+log(S/H)                     &6.33$\pm$0.05       & 6.11$\pm$0.02\\
log(S/O)                        &-1.70$\pm$0.08       & -1.72$\pm$0.11\\
Ne$^{++}$/H$^{+} \times$10$^{5}$&1.55$\pm$0.16       & 0.92$\pm$0.10 \\
ICF(Ne)                         &1.65$\pm$0.13       & 1.10$\pm$0.21 \\
Ne/H $\times$10$^{5}$           &2.55$\pm$0.50       & 1.01$\pm$0.32\\
12+log(Ne/H)                    &7.41$\pm$0.19       & 7.01$\pm$0.31\\
log(Ne/O)                       &-0.62$\pm$0.23   & -0.82$\pm$0.41\\
\hline
He II $\lambda$4686/H$\beta$                             & 0.008$\pm$0.002 & 0.016$\pm$0.005\\
log($\left[O III\right]\lambda$5007/H$\beta$)            &   0.54$\pm$0.05 & 0.77 $\pm$0.02\\
log($\left[N II\right]\lambda$6584/H$\alpha$)            &  -1.17$\pm$0.09 & -1.97$\pm$0.12\\
log($\left[S II\right]\lambda\lambda$6717,6731/H$\alpha$)&  -0.87$\pm$0.09 & -1.47$\pm$0.16 \\
$\sigma$(H$\alpha$) km s$^{-1}$                          & 44.47           & 25.75\\
Age Myr                                                  & 4.95            &  4.12\\
\hline
\end{tabular}
\label{integrated_properties_1}
\end{minipage}
\end{table}

\textit{Tol 0104-388}. 
In this galaxy, the electron density is relatively high (upper panel of
Fig. \ref{ratios_den}), varying from $\sim$100 to 1033 cm$^{-3}$, 
while  the temperature varies from 10502 K to 13700 K. 
The highest values of density are placed near the peak of H$\alpha$ emission.
We obtained an integrated electron temperature of T$_{e}$(O \,{\sc iii})= 12199$\pm$412 K, that agree with the
value reported by \citet{P91} of 1.25$\pm$0.15 $\times$10$^{4}$ K, and a density 
of n$_{e}$(S \,{\sc ii})=614 cm$^{-3}$. 
The oxygen abundance values in units of 12+log(O/H) in Tol 0104-388 range 
from 7.96 to 8.24, with a mean value of 8.06 and a standard deviation of 0.06. 
We obtained an integrated value of 12+log(O/H)=8.02$\pm$0.04 that agrees with the mean value of the spaxels within the errors.
This integrated 12+log(O/H) abundance is consistent with the value of 12+log(O/H)=8.19$\pm$0.15 
derived by \citet{P91}. The N$^{+}$/H$^{+}$ distribution range values from 1.85 $\times$10$^{-6}$ 
to 2.72 $\times$10$^{-6}$ with a mean value of 2.21 $\times$10$^{-6}$. 
With this information we found that the 12+log(N/H) range values from 6.60 to 6.92, a mean 
value of 6.74, a standard deviation of 0.07 and an integrated value 
of 12+log(N/H)=6.77$\pm$0.08. Finally, the log(N/O) ratio range values from -1.43 to -1.20 
with a mean value of -1.33 and a standard deviation of 0.05.
We found an integrated value of \mbox{log(N/O)=-1.25$\pm$0.12}, \mbox{log(Ne/O)=-0.62$\pm$0.23} 
and \mbox{log(S/O)=-1.70$\pm$0.08}. These values appear normal for BCD galaxies at this metallicity \citep{I99,LS10a,G11}.

\textit{Tol 2146-391}. In this galaxy,
we obtained an integrated temperature of T$_{e}$(O \,{\sc iii})=15277$\pm$584 K 
and a density of n$_{e}$(S \,{\sc ii})$\la$100 cm$^{-3}$. 
The electron temperature varies from 12802 K to 22525 K. 
The density in the  outer part of the galaxy
is practically constant with n$_{e}$(S \,{\sc ii})$\la$100 cm$^{-3}$, given that the ratio 
[S \,{\sc ii}]$\lambda$6717/[S \,{\sc ii}]$\lambda$6731 is typically 
greater than 1 (see Fig. \ref{ratios_den}) which indicates a low density regime \citep{O06}, hence, we assumed an 
electron density of n$_{e}\sim$100 cm$^{-3}$ in most apertures in Tol 2146-391. 
Our n$_{e}$ values found in the inner part of the galaxy are consistent with the 
values found by \cite{G11} of 162 cm$^{-3}$ and 147 cm$^{-3}$, corresponding with the two GH\,{\sc ii}Rs (Fig. \ref{image}) found
in this work.  
The oxygen abundance 12+log(O/H) range from 7.69 
to 7.96, with a mean value of 7.84 and a standard deviation of 0.06. 
The integrated oxygen abundance of 12+log(O/H)=7.82$\pm$0.09 agree with the value  
obtained by \citet{P06} of 7.80$\pm$0.01, 7.78$\pm$0.01 by \cite{G07} and 7.82$\pm$0.01-7.79$\pm$0.01 (for the two GH\,{\sc ii}Rs) 
by \cite{G11} but is in disagreement with the reported value of 7.62$\pm$0.08 given by \citet{K06}.
The N$^{+}$/H$^{+}$ distribution range values from 0.19 $\times$10$^{-6}$ to 0.44 $\times$10$^{-6}$ with a mean 
value of 0.29 $\times$10$^{-6}$ and a standard deviation of 0.06 $\times$10$^{-6}$.
Given that our calculated ICF(N), over the majority of the spaxels in the FoV, are typically $\sim$10 
we found a variation of 12+log(N/H) from 6.27 to 6.83,
an average value of 6.52 dex and a standard deviation of 0.16.
An integrated value of 12+log(N/H)=6.54$\pm$0.07 is found in Tol 2146-391. 
The log(N/O) ratio range values from -1.55 to -1.05 
with an average value of -1.32 and a standard deviation of 0.12.
We found a value of log(N/O)=-1.28$\pm$0.29 summing over all spaxels in the FoV. 
This integrated value appear to be in agreement, within the errors, with the ones obtained by
\cite{G11} for the two GH\,{\sc ii}Rs log(N/O)=-1.57$\pm$0.03 and -1.63$\pm$0.03.
Finally, our integrated log (Ne/O)=-0.82$\pm$0.41 appear to be consistent with values obtained by \cite{G11}.

\begin{figure}
\centering
\includegraphics[width=64mm]{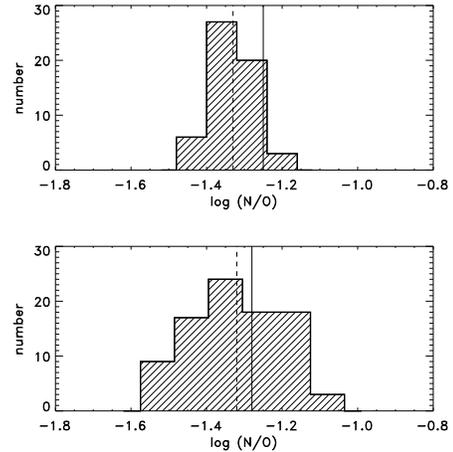}
 \caption{\textbf{Distribution of log (N/O), in Tol 0104-388 (upper panel) and Tol 2146-391 (lower panel)}.
For each of these two galaxies we show the integrated log (N/O) values (solid lines) 
and the mean log (N/O) values (dotted lines).
}
\label{NO_distri}
\end{figure}

In Fig. \ref{NO_distri} we show
the distribution of log (N/O), in Tol 0104-388 and Tol 2146-391, respectively.
For each one of these two galaxies we show the integrated log (N/O) values (solid lines) 
and the mean log (N/O) values (dotted lines) of the data points.
Table \ref{static} summarize the statistical properties of the spaxels
in our sample galaxies. We note an agreement, within the errors, between the mean value of the spaxels 
and the properties derived from the integrated spectrum. 
In the case of the oxygen abundance, we found a variation of $\Delta$(O/H)$\sim$0.28 between
the minimum and maximum values, for both galaxies, with a low standard deviation.
This result indicates that these variations are not statistically significant  
across the galaxies. The log (N/O) variation in Tol 0104-388
shows a value of 0.22 with a low standard deviation.
While, in Tol 2146-391 we found a variation of $\Delta$(N/O)=0.50, with a higher standard
deviation with respect to the oxygen abundance. Although the mean value of log (N/O) is consistent 
with the integrated value, this result indicates that a slight variation is observed in Tol 2146-391. 
In fact, Fig. \ref{abundance_maps} and \ref{NO_maps} show that the N$^{+}$ ions and consequently
12+log(N/H) show a slightly positive trend from the region near the H$\alpha$ peak 
to the North-West part of the galaxy, producing the increasing of the N/O ratio to this direction, 
reaching values of log(N/O) $\ga$-1.20.

\begin{table*}
 \centering
 \begin{minipage}{100mm}
  \caption{Statistical properties of the galaxies.}
  \begin{tabular}{@{}lccccclrlr@{}}
  \hline      
&    \multicolumn{3}{c}{Tol 0104-388}& \multicolumn{3}{c}{Tol 2146-391}\\
& Mean & SDEV\footnote{Standard deviation of the spaxels.} & $\mid\Delta$\footnote{Difference between minimum and maximum values.}$\mid$& Mean 
& SDEV & $\mid\Delta\mid$\\

 \hline
O$^{+}$/H$^{+}$ ($\times$ 10$^{5}$) &4.73  &0.63 &2.91 &0.60 &0.11 &0.39\\
O$^{++}$/H$^{+}$ ($\times$ 10$^{5}$)&7.12  &1.24 &5.68 &6.34 &0.99 &4.37\\
12 + log (O/H)                      &8.06  &0.06 &0.28 &7.84 &0.06 &0.27\\
N$^{+}$/H$^{+}$ ($\times$ 10$^{6}$) &2.21  &0.00 &0.87 &0.29 &0.06 &0.25\\
12 + log (N/H)                      &6.74  &0.07 &0.31 &6.52 &0.16 &0.56\\
log (N/O)                           &-1.33 &0.05 &0.22 &-1.32&0.12 &0.50\\ 
\hline
\end{tabular}
\label{static}
\end{minipage}
\end{table*}

\section{Discussion}\label{discussion}
 
\subsection{Spatial distribution and Dependence of the He \,{\sc ii} $\lambda$4686 emission line 
on the EW(H$\beta$) and metallicity of the ISM}\label{ew_met}

\begin{figure}
\centering
\includegraphics[width=50mm]{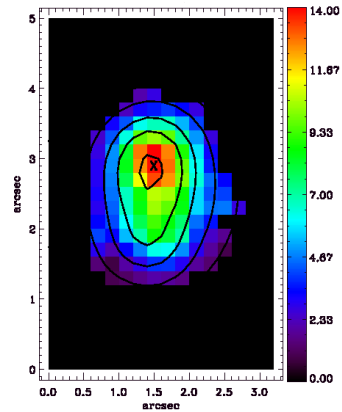}
 \caption{
Spatial distribution of the He \,{\sc ii} $\lambda$4686 emission line 
in the galaxy Tol 2146-391. Fluxes in units of  10$^{-18}$ ergs cm$^{-2}$ s$^{-1}$ $\rm \AA^{-1}$. 
The maximum H$\alpha$ emission is indicated in the map by an X symbol.
}
\label{high-ioni-Tol21}
\end{figure}

\begin{figure}
\centering
\includegraphics[width=90mm]{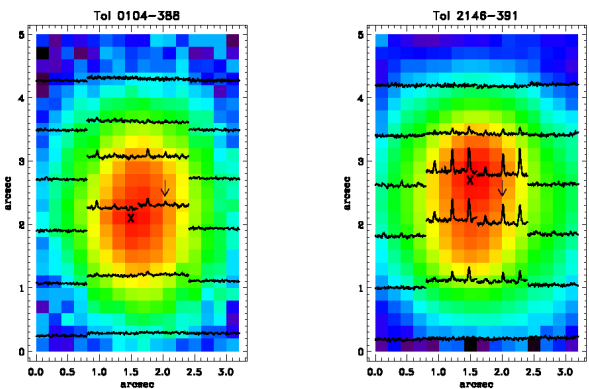}
 \caption{
Spatial distributions of the He \,{\sc ii} $\lambda$4686 emission line  
in Tol 0104-388 and Tol 2146-391. In this figure we show the H$\beta$ emission line map in the original 
0\arcsec.2 pixel size and superimposed the binned 0\arcsec.8 spaxels in the region that 
contain the He \,{\sc ii} $\lambda$4686 emission line.  
The maximum H$\alpha$ emission is indicated in the maps by an X symbol. 
The arrows in the Figures are indicating the position
of He \,{\sc ii}  $\lambda$4686 emission lines in the spectra.
}
\label{high-ioni}
\end{figure}

In Fig. \ref{high-ioni-Tol21} we show the spatial distribution of He \,{\sc ii} $\lambda$4686
in Tol 2146-391. Fig. \ref{high-ioni} shows the spatial distribution 
of H$\beta$ emission lines in the 0\arcsec.2 pixel scale, for both galaxies, and superimposed in each one of the images 
the resampled 0\arcsec.8 spaxels in the region that contain the He \,{\sc ii} $\lambda$4686 emission line 
from 4640 $\rm \AA$ to 4725 $\rm \AA$. To do this we summed the spaxels in a region of 4 $\times$ 4 spaxels in order to increase 
the S/N. From this fig., we find a weak He \,{\sc ii} $\lambda$4686 emission line in the core of Tol 0104-388, previously not
detected in this galaxy, so we see that in both galaxies the He \,{\sc ii} $\lambda$4686 emission line is concentrated 
in the core. 

\citet{TI05} studying a sample of galaxies where the He \,{\sc ii} $\lambda$4686 emission line
was detected, found that the hardness of the ionizing radiation does not appear to depend 
on starburst age or EW(H$\beta$), when the integrated properties
of the galaxies are considered. On the other hand,  
as mentioned by previous spectroscopic studies \citep[e.g.,][]{C86,G00,TI05}
the hardness of this ionizing radiation in BCDs increases with decreasing metallicity.  
This implies that these ionization emission lines are stronger in galaxies 
with lower metallicities. 
This correlation was determined from the integrate spectra of galaxies. 
But what is the relationship between the spatial distribution of abundances and EW(H$\beta$) 
with the hardness of high ionizing radiation across the spaxels in our sample galaxies? 

\begin{figure*}
\includegraphics[width=150mm]{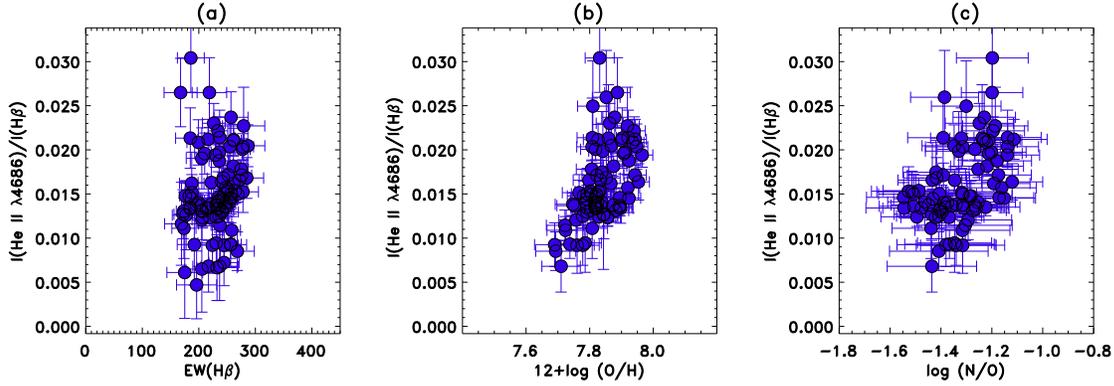}
 \caption{
Intensity of the He II $\lambda$4686 emission line relative to H$\beta$ as a function 
of the equivalent width of H$\beta$, oxygen abundance 12 + log (O/H) and log (N/O) ratio
for the galaxy Tol 2146-391.
}
\label{EW_OH_HeII}
\end{figure*}

In Fig. \ref{EW_OH_HeII} we show the intensity of the He \,{\sc ii} $\lambda$4686 emission line
relative to H$\beta$ of each spaxel of 0\arcsec.2 as a function of (a) EW(H$\beta$), 
(b) oxygen abundance 12 + log (O/H) and (c) log (N/O) ratio for the galaxy Tol 2146-391. 
We performed a Spearman's rank correlation test on our data in order to assess how well 
the relationship between He \,{\sc ii} $\lambda$4686 emission line 
can be described as a function of the other properties. 
From this procedure, we did not find a significant correlation between these variables using
this correlation technique. 
In fact, the oxygen abundances appear to be uniform
in the regions where the He \,{\sc ii} $\lambda$4686 emission line was measured. 
It can be interpreted that there are no correlations with small patches of the ISM. 

\subsection{Sources of the nebular He \,{\sc ii} $\lambda$4686 emission line 
and their spatial distribution}\label{high}

Studing a sample of GH\,{\sc ii}Rs, \citet{G00} found that galaxies
with detected and nondetected WR features have the same distribution
dependences of I(He \,{\sc ii} $\lambda$4686)/I(H$\beta$) versus EW(H$\beta$). This implies that WR stars are not
the sole origen of He \,{\sc ii} $\lambda$4686 in star-forming regions.
On the other hand, \cite{DM98} found a spatial correlation between the nebular He \,{\sc ii} $\lambda$4686 emission between
the star clusters and the position of WR stars in the galaxy IZw 18 supporting the hypothesis that WR stars are responsible 
for nebular He \,{\sc ii} emission in this galaxy. While, \cite{I06a} found that the hard ionizing radiation 
responsible for the He \,{\sc ii} $\lambda$4686 emission, in the galaxy SBS 0335-052, is not likely related with the position of the most 
massive and young star clusters, but rather is related to fast radiative shocks. 
Another possibility is that the high-ionizing radiation is produced by the accretion of gas by HMXBs \citep{G91}
located in the star cluster population in the core of the galaxies.  
As suggested by \citet{T04} the high X-ray luminosities of these sources may be due to a metallicity effect, 
resulting in a larger X-ray luminosity, so producing an additional photoionization of the gas
in low-metallicity systems, as is the case of H\,{\sc ii}/BCD galaxies.
Finally, \citet{B08} suggest that at low metallicity the main source of He \,{\sc ii} $\lambda$4686
ionizing photons appears to be O stars.

\begin{figure}
\centering
\includegraphics[width=90mm]{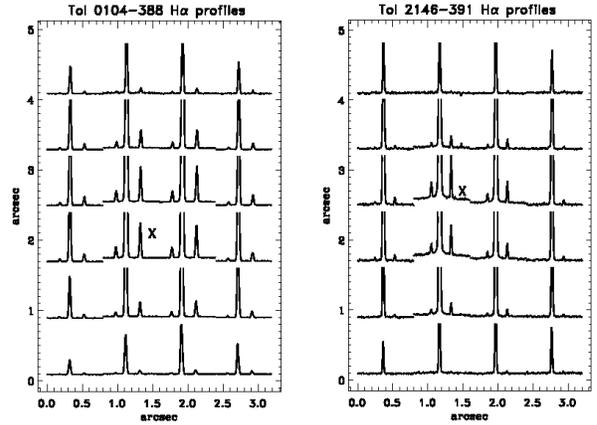}
 \caption{
Spatial distribution of H$\alpha$ profiles in the galaxies Tol 0104-388 and Tol 2146-291.
}
\label{Ha_profiles}
\end{figure}

The action of stellar winds and supernovae (SNe) explosions from the star clusters 
generates expanding shells \citep[e.g.,][]{H74,M98} that eventually will be able to produce on large scales 
collisionally excited emission lines. 
\citet{G91} have suggested that fast radiative shocks in GH\,{\sc ii}Rs can produce 
relatively strong He \,{\sc ii} emission under certain conditions. 
Hydrodynamical models by \citet{DS96} have shown that fast shocks, with shock 
velocities of 400-500 km s$^{-1}$, are an efficient means to produce a strong 
local UV photon field in the ISM of galaxies. 
Therefore, such shocks can be responsible for the observed fluxes of He \,{\sc ii} $\lambda$4686 
and other high ionization emission lines. The existence of these fast motions is supported 
by the presence of broad components, with velocities of hundreds km s$^{-1}$ 
\citep[e.g.,][]{I96,W07}, in the line profiles 
of the strongest emission lines in some HII/BCD galaxies and GH\,{\sc ii}Rs. 
We checked for the presence of broad components in the H$\alpha$ line profile 
of our sample galaxies. In Tol 0104-388 this emission line profile is symmetric and well 
represented by a single Gaussian, and does not show prominent low intensity broad components 
in the integrated spectrum.
In the case of Tol 2146-291 the base of the H$\alpha$ (and other such as [OIII] $\lambda$5007) 
profile appears to be very broad,  
with a component similar to the ones observed in other BCD galaxies \citep[e.g.,][]{I96}. 
Fig. \ref{Ha_profiles} shows the spatial distribution 
of H$\alpha$ emission line profiles in the 0\arcsec.8$\times$0\arcsec.8 spaxel size, 
as indicated in Sect.~\ref{ew_met}, for the galaxies Tol 0104-388 and Tol 2146-291, respectively. 
Again, we observe that the profile in the spaxels in Tol 0104-388 are well represented by a Gaussian,
while in Tol 2146-291 we observe a broad base in the emission lines in the spaxels in the inner part 
of the map (see Fig. \ref{Ha_profiles}). 
In order to detect the presence of a broad component in these spaxels, we used 
the PAN\footnote{http://www.ncnr.nist.gov/staff/dimeo/panweb/pan.html} routine (Peak ANalysis) 
in IDL to fit two components to these profiles. 
Using this software, we found that these spaxels are well represented by a single Gaussian,
with the exception of the spaxel on the peak of H$\alpha$ emission. 
The broad component was estimated to be $\sigma_{broad} <$100 km s$^{-1}$ 
with F(H$\alpha$)$_{broad}$/F(H$\alpha$)$_{total}$ $\sim$ 30\% 
and F(H$\alpha$)$_{total}$=F(H$\alpha$)$_{narrow}$+F(H$\alpha$)$_{broad}$.

Some evidences of the presence of two components in the emission line profile in this region 
are found in \cite{B11}, where they observed a double component in the base of the emission lines  
H$\alpha$, H$\beta$, [O\,{\sc iii}]$\lambda$4959 and [O\,{\sc iii}]$\lambda$5007 in the integrated spectrum 
of Tol 2146-391. We must be careful in the interpretation of our kinematic results given 
that we use a medium resolution and not a high resolution grating. 
In any case, we may speculate that a broad component, eventually, 
could be properly resolved using high resolution spectroscopy in Tol 2146-391 and it may be associated with 
unresolved wind-driven shells \citep[e.g.,][]{C94}. 
This scenario is plausible given that expanding structures 
have been detected in some compact H\,{\sc ii}/BCD galaxies, using H$\alpha$ and/or 
H$\beta$ emission line maps \citep[e.g.,][]{P02,L07}.
We have found in previous studies \citep{L07} these filamentary structures
in galaxies with one and multiple SF regions, such as Tol 0957-278, Tol 1004-296, UM 456, UM 462, UM 463,
UM 483, IIZw 70 and Tol 1924-416. One for instance can note the filamentary structures generally very close 
to the star cluster complexes, to distances that range from $\sim$100 pc to $\sim$1000 pc, while the EW(H$\beta$) map 
of these galaxies shows these structures associated with large EW(H$\beta$) values \citep{L07}.

Finally, as we mentioned previously in this work, only Tol 2146-391 was reported
in the literature as a WR galaxy by previous studies \citep[e.g.,][]{M91}. 
The examination over each individual 0\arcsec.2 spaxel and in the binned maps (see Fig. \ref{high-ioni})
and also in the integrated spectra (see Fig. \ref{integrated_spec}), 
in both galaxies in our sample, do not reveal any clear stellar WR features.  
Given the spatial distribution of the nebular He \,{\sc ii} $\lambda$4686 emission 
in our two analyzed galaxies, this high ionizing radiation is likely associated 
with a mix of sources, where WR stars, HMXB and O stars 
cannot be excluded. 
Expanding shells powered by these unresolved star cluster/complexes \citep{L11} can likely also
be added as a source of ionizing radiation, at least in the case of Tol 2146-391.

\subsection{Spatial and radial correlation of properties in the ISM} 

\begin{figure}
\centering
\includegraphics[width=84mm]{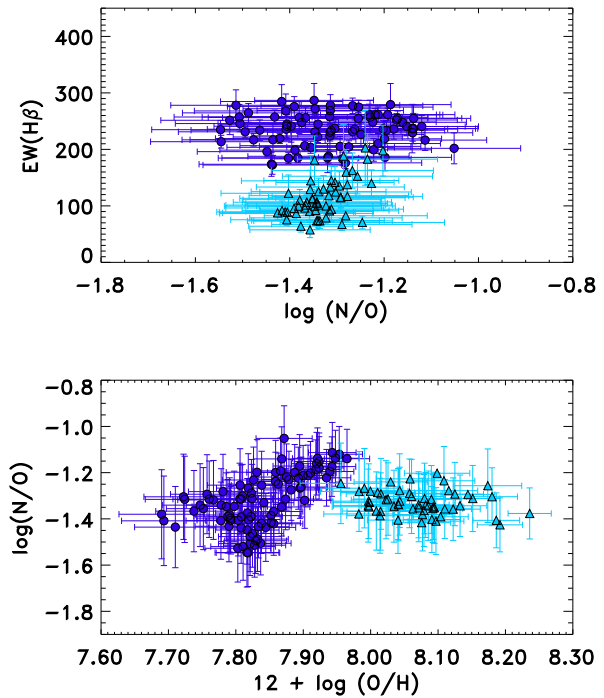}
 \caption{
log (N/O) ratio versus EW(H$\beta$) and 12 + log (O/H) ratio versus log(N/O).
Triangles correspond to the data points of Tol 0104-388 and circles corresponds
with the data points of Tol 2146-391.
}
\label{NO_EW}
\end{figure}

In Fig. \ref{NO_EW} we show the log (N/O) versus EW(H$\beta$) and 12+log(O/H) vs. log (N/O) 
for all 0\arcsec.2 spaxels in our two galaxies. 
In this fig., triangles correspond to the data points 
of Tol 0104-388 and circles correspond to the data points of Tol 2146-391.
\citet{I06b} suggest that there is a dependence between the N/O ratio and the
EW(H$\beta$), for the entire galaxies, in the sense that the N/O ratio should increase
with decreasing EW(H$\beta$). 
Their conclusion is based on integrated spectra of a large sample of emission line galaxies 
with EW(H$\beta$) ranging from $\sim$20 to $\sim$350 $\rm\AA$.
This trend was also observed recently by \cite{B08} and \cite{LS10a}. 
But, no correlation between N/O and EW(H$\beta$) is found by \cite{B05} using a smaller sample of galaxies.
\citet{I06b} argue that the observed trend of N/O increasing as EW(H$\beta$) decreases is naturally explained 
by the expected ejection from WR stars. 
We observed in the upper panel of Fig. \ref{NO_EW} that the EW(H$\beta$) values are rather constant,
with a very small range of equivalent widths $\Delta$EW(H$\beta$) $\sim$100$\rm \AA$, when
the N/O ratio increases. Therefore, no correlation between EW(H$\beta$) and N/O is found in our sample galaxies.
The reader must bear in mind that, in our case, we are concerned with the distribution of the physical properties 
within an individual galaxy and how that may affect the internal physical processes, as opposed to the statistical 
trends among different galaxies.
If we compare the log (N/O) vs. 12+log(O/H), in the lower panel of Fig. \ref{NO_EW} for Tol 2146-391,
we see that the log (N/O) values increase with the 12+log(O/H) abundance from $\sim$7.70
to $\sim$7.97. This data point distribution has similar patterns to those found in BCD/HII galaxies \citep[e.g.,][]{I99}  
of increasing N/O ratios with respect to the oxygen abundance. These studies suggest that at low metallicity (12+log(O/H)$<$7.6)
the N is associated with primary processes, at intermediate metallicities with a combination of primary and secondary processes, while
at higher metallicities (12+log(O/N)$\ga$8.30) only with secondary processes. 
The inner region of Tol 2146-391 (near the peak of H$\alpha$) present N/O
ratios which are larger than those expected by the pure primary
nature of nitrogen. This might be a signature of time delay between
the release of oxygen and nitrogen \citep{KS98}, or even the presence
of dynamical processes such as gas infall or outflow.
In any case, for the metallicity of Tol 2146-391 secondary processes are unplausible.
No correlation is found between log(N/O) vs. 12+log(O/H) in Tol 0104-388. 

\begin{figure*}
\includegraphics[width=120mm]{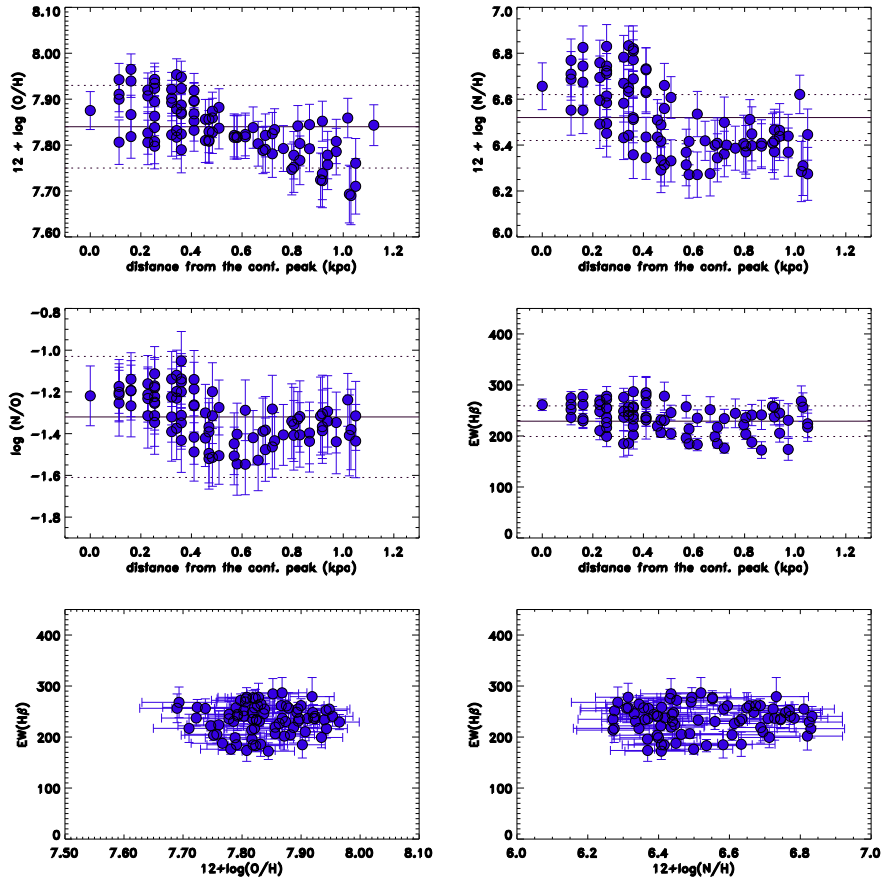}
 \caption{
12+log (O/H), 12+log(N/H), log (N/O) ratio and EW(H$\beta$) as a function of
the distance from the H$\alpha$ continuum peak and 12+log (O/H); and 12+log(N/H) versus EW(H$\beta$)
for the galaxy Tol 2146-391. 
The integrated properties are indicated with a black continuous line. The
two dotted lines shows the uncertainties of these values.
}
\label{kpc-properties}
\end{figure*}

In Fig. \ref{kpc-properties} we show the radial distribution of oxygen abundance, nitrogen abundance, log(N/O),
EW(H$\beta$) with respect to the H$\alpha$ continuum peak and EW(H$\beta$) with respect to 12+log(O/H) and 12+log(N/H)
for the galaxy Tol 2146-391.
The integrated properties of the galaxy are indicated with a continuous line and the two dotted lines show the uncertainties of
these values at the 1$\sigma$ level. Statistically, the bulk of data points lie in a region of $\pm$1$\sigma$ 
around the integrated value of oxygen and log(N/O) radial distribution. 
These results indicate that there is no significant variation across the galaxy. 
However, the 12+log(N/H) radial distribution shows that the highest values are located near the peak of continuum emission, 
with values reducing with radius.

If real, the observed trend of increasing 12+log(N/H) abundance in Tol 2146-391 would argue in favour 
of self-enrichment by the fresh heavy elements during the present burst of SF, on scales of hundreds of pc, or
alternatively these heavy elements were produced during the previous burst of SF and dispersed in large
scales across the galaxy in the current burst. 
If we assume that an instantaneous burst is a better description of the current burst in Tol 2146-391, 
we can obtain, using STARBURST99,
that the mechanical luminosity released into the ISM via radiative winds and
SNe in order to produce a super-shell is $\sim$12$\times$10$^{39}$ erg s$^{-1}$. The radius in pc of an expanding super-shell 
can be written as R$_{s}$=269(L$_{38}$/n$_{0}$)$^{1/5}$t$_{7}^{3/5}$ \citep{MK87}, where L$_{38}$ denotes the mechanical luminosity in units 
of L$_{38}$ erg s$^{-1}$, t$_7$ = t/10$^7$ yr and n$_0$ is the density of the ambient gas in cm$^{-3}$. So,
the distance reached by the super-shell at 4 Myr is R$_{s}\sim$0.6 kpc with 
a velocity of v$_{s}\sim$241 km s$^{-1}$, assuming n$_0\sim$100 cm$^{-3}$. 
Hence, the expansion of these shells is sufficient for producing, within the age of the current burst, the dispersion 
of the metal at a distance R$_{s}\la$0.4 kpc. 
On the other hand, \cite{P12} determined the physical conditions 
in Tol 2146-391 considering the presence of thermal inhomogeneities (t$^{2}$). The high value of t$^{2}$ found by Peimbert et al.,
in Tol 2146-391, agree with the idea that radiative shocks coming from SNe explosions have higher
effect on the thermal structure of the galaxy.
In any case, a conclusive assessment of this issue is not possible with the present data, 
but this scenario appears to be viable and agrees within the overall observed properties of Tol 2146-391.

As we mentioned previously in Sect.~\ref{high}, the examination of each individual 0\arcsec.2 spaxel 
and in the binned maps (see Fig. \ref{high-ioni}), in both galaxies in our sample, does not reveal any clear WR feature.  
The detection of WR stars is not easy and different studies show different results. 
For example in the galaxy UM 420 \citet{LS10b}, using long slit spectroscopy, detected the presence of WR
stars, but using VLT-VIMOS observations \citet{J10} did not find any evidence of WR signature in this galaxy. 
This discrepancy indicates that the detection of WR features in galaxies depends on the quality of the spectra, 
location and size of the apertures \citep{S99}. Therefore, if the detection of WR stars in Tol 2146-391 is not proven, 
the observed trend could be due to the outflow of previously enriched gas. 
Since the metallicity and composition of the ISM evolves with time, the super-shell wind
will produce a metallicity gradient, with the inner parts containing a larger proportion
of metals. In the case of Tol 2146-391, the presence of these energetic winds are supported by the observation 
of an irregular and broad component in the base of the emission lines \citep{B11}.  

In any case, the results obtained in this work suggest that even though a slight gradient, 
in scales of hundreds of pc, of N is observed in the ISM of Tol 2146-391 
(as result of the starburst activity) the properties across the galaxies are fairly well mixed. 
This suggests that the O and N enrichment and dispersion is likely related with a global 
process in the galaxies. 
\cite{P11} studied a sample of three BCD galaxies with high value of N. 
They found that the globally high N/O ratio in these objects is not likely produced by stellar winds 
coming from WR stars, but on the contrary, could be related more to other global processes
affecting the metal content of the whole galaxy.  
We suggest that global hydro-dynamical effects, such as starburst-driven super-shells 
might be attributed to efficient transport and mixing of metals across the galaxies,
so keeping the N/O ratio constant through the ISM at large scales.
These no significant variation in abundance across the ISM of the galaxies would be indicative of uniform SF history,
that is likely the case of low luminosity and compact H\,{\sc ii} galaxies, where the SF appear to be simultaneous 
over short timescales \citep{L11}. 
This scenario agrees with the idea that most of the He \,{\sc ii} $\lambda$4686 emission is
produced by radiative shocks powered by a plethora of unresolved star clusters.
Therefore, these H\,{\sc ii} galaxies are a genuine example of the simplest starburst occurring in galactic scale, 
possibly mimicking the properties one expects for young galaxies and/or H\,{\sc ii} galaxies at intermediate 
and high redshift.

\section{Conclusions}

Using new GMOS--IFU spectroscopic observations of the compact 
H\,{\sc ii}/BCD galaxies Tol 0104-388 and Tol 2146-391, we studied 
the spatial distribution of the high-ionization emission line 
He \,{\sc ii} $\lambda$4686 and the chemical pattern through the
ISM of the galaxies in an extended region of 3$\arcsec$.2$\times$5$\arcsec$, 
equivalent to $\sim$1372 pc $\times$ 2058 pc and 
$\sim$1820 pc $\times$ 2730 pc for Tol 0104-388 and Tol2146-391, respectively. 
Based on the analysis of its properties, we have obtained the following results:

\begin{enumerate}
\item The examination over each individual 0\arcsec.2 spaxel
and also in the integrated spectra, 
in both galaxies in our sample, do not reveal any clear stellar WR features. 

\item Both galaxies show the presence of the emission line 
He \,{\sc ii} $\lambda$4686 with an integrated intensity relative to H$\beta$ 
of I(He \,{\sc ii} $\lambda$4686)/I(H$\beta)<$0.02. 
We did not detect a clear correlation between the spatial distribution 
of EW(H$\beta$), 12 + log (O/H) and log(N/O) with respect to the hardness 
of this high-ionization radiation across the spaxels in Tol 2146-391. 

\item Given the spatial distribution of He \,{\sc ii} $\lambda$4686 emission in 
our two analyzed galaxies, this high ionizing radiation is likely 
associated with a mix of sources, where WR stars, HMXB and O stars cannot be excluded. 
While, expanding shells powered by a plethora of unresolved star clusters
are likely producing most of the observed 
He \,{\sc ii} $\lambda$4686 emission in our sample galaxies, at least in Tol 2146-391. 

\item We found some evidence that the 12+log(N/H) radial distribution, in Tol 2146-391, shows a slight trend,
with the values decreasing with distance from H$\alpha$ continuum peak.
If real, this observed trend of 12+log(N/H) abundance
would argue in favour that these heavy elements were produced during the previous burst of SF and are currently dispersed 
by the expansion in the ISM of starburst-driven super-shells.
However, the spatial constancy of the N/O ratio might be attributed to efficient transport 
and mixing of metals by hydro-dynamical effects during the previous episodes of SF. 
\end{enumerate}

All results presented here are suggestive that the physical conditions in these two galaxies, as in the case 
of the low luminosity and compact galaxy UM 408 \citep{L09}, vary in a very small dynamical range and are quite homogeneus. 
Therefore, the lack of significant variation in abundance across the ISM of the galaxies would be indicative of uniform SF history 
occurring in galactic scales.

\section*{Acknowledgments}

We would like thank the anonymous referee for his/her comments and suggestions which substantially improved the paper.
P.L. is supported by a Post-Doctoral grant SFRH/BPD/72308/2010, 
funded by FCT (Portugal). 
P.L. would like thank to Polychronis Papaderos and Andrew Humphrey for their very useful comments, suggestions and  
discussions which have improved the paper.
Based on observations obtained at the Gemini Observatory, which 
is operated by the Association of Universities for Research in Astronomy, Inc., under a cooperative agreement 
with the NSF on behalf of the Gemini partnership: the National Science Foundation (United States), the Science 
and Technology Facilities Council (United Kingdom), the National Research Council (Canada), CONICYT (Chile), 
the Australian Research Council (Australia), Minist\'erio da Ci\^encia e Tecnologia (Brazil) and Ministerio de Ciencia,
Tecnolog\'{\i}a e Innovaci\'on Productiva (Argentina). Gemini Program ID: GS-2004B-Q-59 and GS-2005B-Q-19.

\label{lastpage}

\end{document}